\documentclass{sf2a-conf2012}
\usepackage{graphicx}
\usepackage{hyperref}
\usepackage[]{natbib}  
\usepackage[cyr]{aeguill}
\usepackage{epstopdf}

\def\BibTeX{{\rm B\kern-.05em{\sc i\kern-.025em b}\kern-.08em
    T\kern-.1667em\lower.7ex\hbox{E}\kern-.125emX}}
\bibpunct{(}{)}{;}{a}{}{,}  

\begin{document}

\TitreGlobal{SF2A 2012}


\title{The future of astronomy PhDs in France}

\runningtitle{The future of astronomy PhDs in France}

\author{S. Boissier}\address{Aix Marseille Universit\'e, CNRS, LAM (Laboratoire d'Astrophysique de Marseille) UMR 7326, 13388, Marseille, France} 

\setcounter{page}{237}


\maketitle


\begin{abstract}
This contribution presents a poll undertaken at the beginning 
of 2012, and addressed to every doctor in 
astronomy who obtained his/her degree in France. Its goal is to motivate
the French astronomical community to think and discuss about what 
should be the training of PhDs, and what should be its objective. 
Further discussions and reactions can be posted e.g. on \url{http://docastro.blogspot.fr/}.
A worrying results from the poll
is that the majority of the participants would not encourage a
young student to start a thesis in astronomy. 
The main reasons for this fact may be the high pressure on 
astronomy positions and the
little interest a doctorate has for other careers in France. I suggest we either
have to modify our formations or reduce the number of thesis starting
each year in astronomy.
\end{abstract}



\section{Introduction and a personal motivation}

\begin{figure}[ht!]
 \centering
 \includegraphics[width=1.\textwidth,clip]{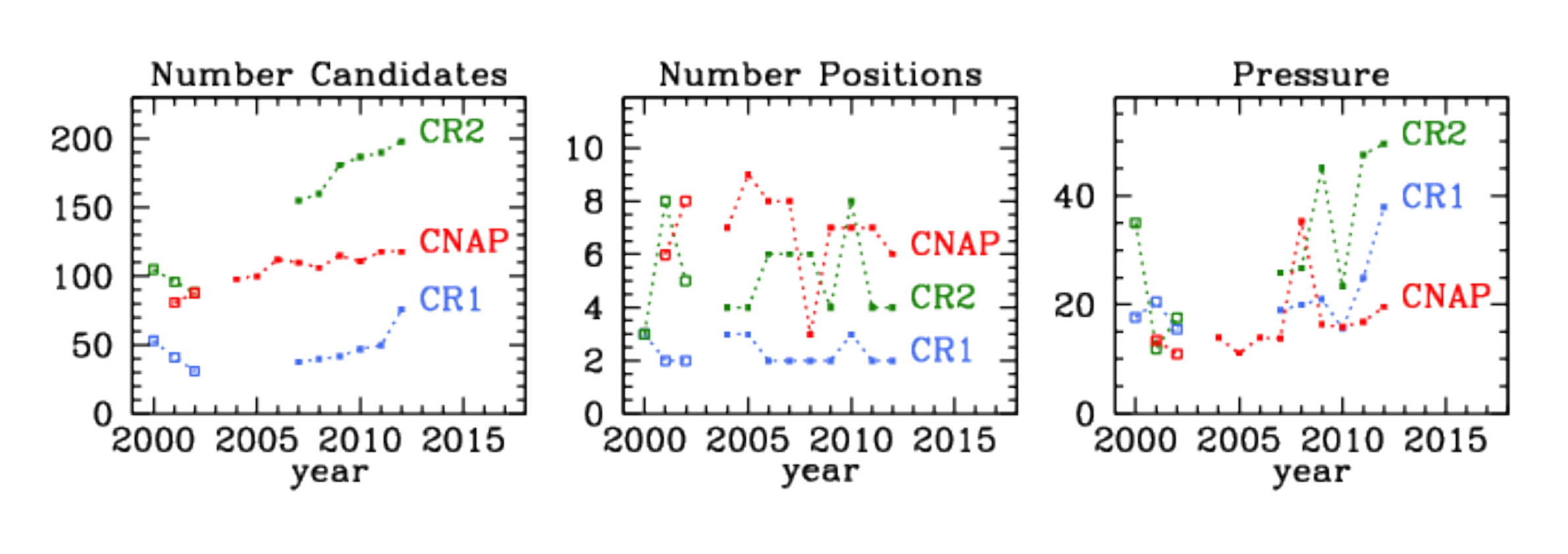}     
 \vspace*{-1cm}
  \caption{Evolution of the number of candidates (left); of the number of 
opened positions at the
national level by the CNRS as CR1 and CR2, and by the CNAP (middle); and 
of the pressure (right). Data compiled by Mederic Boquien 
(private communication) are shown as filled symbols and by Mamon (2003) as open symbols.}
  \label{fig:pression}
\end{figure}

French permanent astronomers are encouraged to take students under
their supervision (enhanced chances of promotions and grants in addition to the
obvious human and scientific interest, as well as 
pressure from pairs: e.g. the astronomy CNRS 
section\footnote{Committee appointed for 4 years 
in charge of both careers evaluation and hiring at 
national positions.}
encourages students supervision when evaluating career progresses). With 
various funding schemes coming into
existence (ERC, ANR), the number of astronomy-related 
thesis has increased during the last 10 years or so. However, the number of 
permanent positions has stayed stable during the same period, leading to an
increasing pressure (see Fig. \ref{fig:pression}). The pressure is even stronger as opportunities in
other countries are becoming scarcer and foreigners are attracted by the French positions.
%
As the number of post-doctoral positions in astronomy has also increased 
(especially via the ANR and ERC schemes), many young doctors stay longer (up to about 10 years, see Fig. \ref{fig:duree}) on a job market offering less and less opportunities. 
Clearly, we are training more and more people in astronomy, 
while many of them will not join the 
astronomy community in the end. What is the meaning of \emph{formation} in this situation ? 

Some astronomers reply that we are performing a ``formation par la recherche'' 
(training through research: a thesis would allow to gain
useful experience for jobs elsewhere than in academia).
However, if in some other countries,
being a ``doctor'' has a meaning in 
society
and is recognized by the private sector, it is 
felt in France as a handicap as enterprises tend to prefer engineers or peopled formed
in our ``grandes \'ecoles'' since the best students are usually drained to them by the
french elitist educational system.
This is despite the fact that doctors are likely to have useful competences to
bring to enterprises (Durette et al., 2012). 
Another ``problem'' is that most students 
starting a PhD in astronomy hope to get a position in research (e.g. Seth et al., 
2009). Said in another way, students do not share the ``formation par la 
recherche'' point of view. They should know their chances of success to end up
in research before starting a thesis (and even a master) in astronomy.

Facing this situation, we cannot supervise students without 
wondering about the interest of long formation in astronomy and
the number of people that should be trained as astronomers.
For this reason,
I started with the help of many people (see acknowledgments) a poll to know what has become of 
french astronomy doctors and how they feel about their current and past experience. 

\section{Other studies}

Before describing this study, I should mention that other people have tackled 
similar questions in the past. All of these works are not as well known
as they should, and I am happy to mention a few initiatives here.

The ``Ecole doctorale d'ile de France'' (concerning only astronomy, what makes their results 
especially pertinent to our community) maintains a site with statistics on the becoming of their
students (\url{http://ufe.obspm.fr/rubrique293.html}). According to them,
about 60 \% of the doctors who got their degree around the year 2000 are 
in research (40\% in France as searchers or engineers, 20\% as searchers abroad).
These numbers are elevated with respect to the situation in other ``Ecoles 
doctorales\footnote{Administrative part of universities in charge of thesis}'', 
at least as described by some of the poll participants.

Turning to astronomy laboratories, 50 \% of people passing their thesis in 
the ``Institut d'Astrophysique de Paris'' (IAP) found a permanent job as a searcher 
in France or abroad (Gary Mamon, private communication).
The ``Institut d'Astrophysique Spatiale'' (IAS) also has a clear record of
the future of their PhD
(\url{https://www.ias.u-psud.fr/website/modules/content_the/index.php?id=9})
showing that about 40 \% get a research or IR (engineer in research)
permanent position in French public jobs for the period
2001-2005. Their results are \emph{posted} on their web-page.
I can only encourage other laboratories to follow this example as this
sort of transparency is \emph{asked} by young people, and necessary if
we assume what we do.

The present study (and those quoted above) concern thesis defended in
France. It is important to take into account the national specificity
for such works. For instance, we should take into account the chance
to have strong national corps in France (CNRS and CNAP).
On the other hand, in e.g. the USA or UK, being a doctor has quite
some valor outside academia, what is not as true in France.  Other
studies may certainly exist but I will quote only a few here. In the
USA, Seth et al. (2009) shows that 170 thesis start per year for 60 to
90 permanent jobs in astronomy (35 to 55 percent of astronomy PhDs
would get permanent positions) without much temporal evolution
(contrary to the French recent increase of the PhDs/jobs ratio).  These
numbers are confirmed by Metcalfe (2007) who mentions that at best 50\% of
astronomy PhDs find a faculty position, while the other half is likely
to end up in ``research and support'' positions.
In comparison, about 50 thesis started each year in Paris during the
last years. It is hard to have complete data, but from what could
be found on internet through a quick search, 7 thesis
started in  Marseilles and 7 more in  Toulouse in the last years on average.
That's already about 75 thesis per year (certainly a lower limit 
since I did not count other important laboratories in 
Strasbourg, Bordeaux, Grenoble, ...) for a total number of jobs each year
(including university, CEA, CNRS and CNAP positions) of about 25.
The comparison of the French and American numbers seems to indicate
better perspectives for young people in the USA.
Nevertheless, Seth et al. (2009) make political recommendations, such
as \emph{``shifting priorities from early
  career temporary positions towards more long-term employment will
  create a more sustainable, equitable, and productive astronomy
  workforce''}.
The French astronomical community should do as much as the
American one and get a critical look at the formation of 
astronomy PhDs in France, and their becoming.  A good analysis of the 
hiring situation in France almost 10 years ago was
presented in Mamon (2003).  To this, should be added the reports from
the CNAP and CNRS sections
that are made at the end of each mandate.
However these concern mostly their own work and we are still missing
clear statistics on the rate of success for PhDs in astronomy to get
permanent positions.
Independent assessment of the situation are too rare, and this paper
is an attempt to discuss the situation and open the debate in the
French community. Everyone is invited to react, for instance 
on the ``docteurs en astrophysique'' (\url{http://docastro.blogspot.fr/}) 
blog. Hopefully, this study could also be useful abroad for
comparison with foreign policies and situations.

\section{Poll description}

\begin{figure}[ht!]
 \centering
 \includegraphics[width=0.45\textwidth,clip]{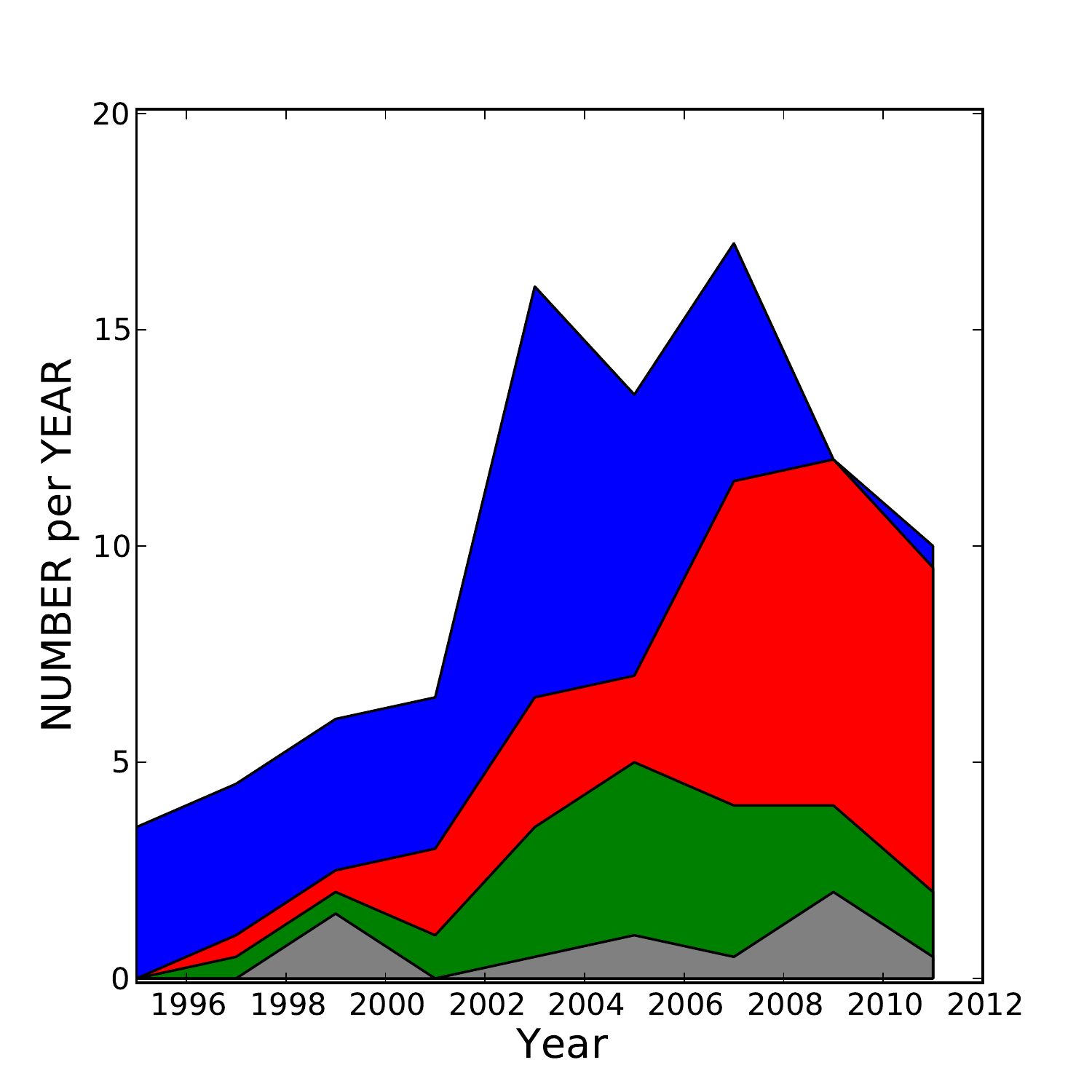}  
 \includegraphics[width=0.45\textwidth,clip]{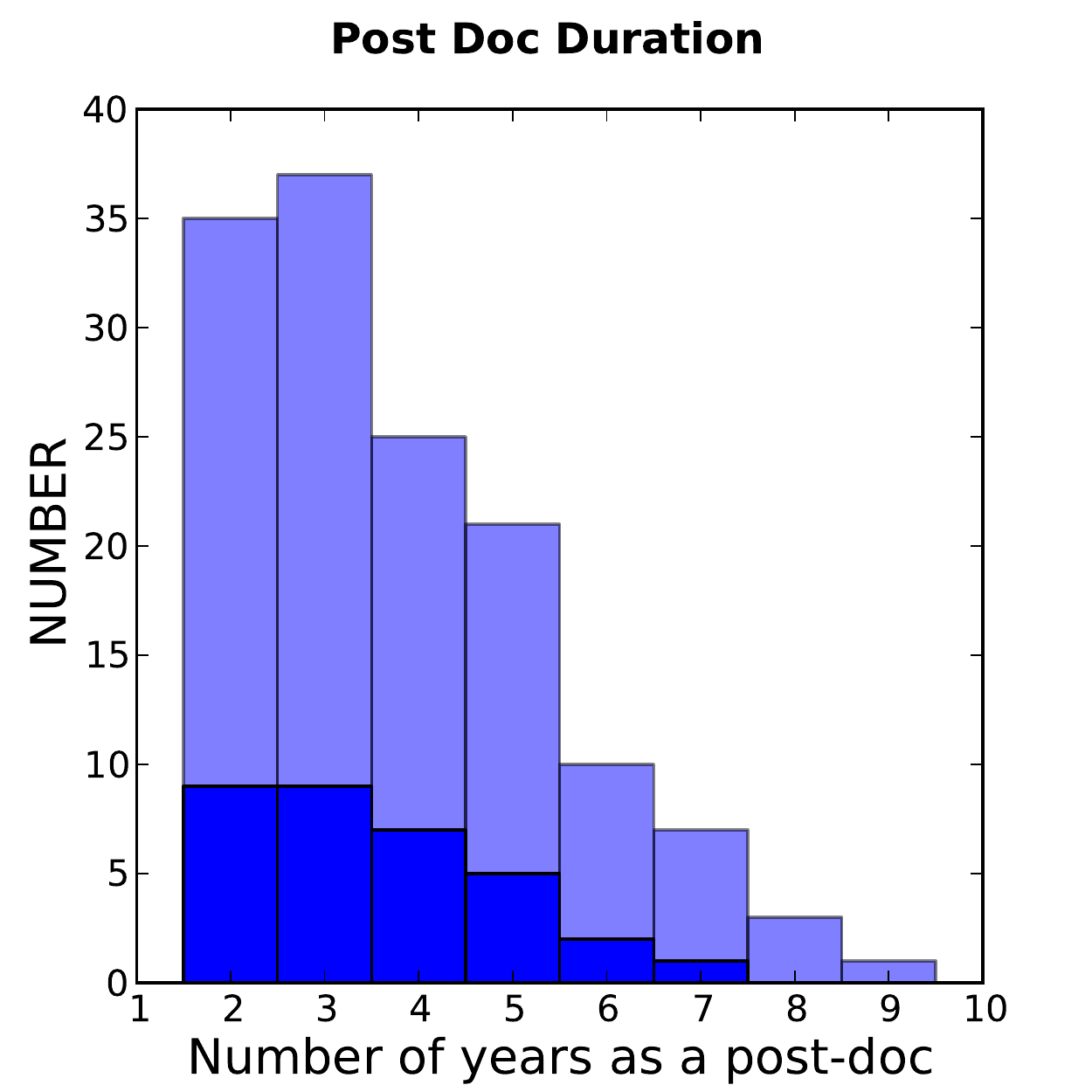}
  \caption{Left: cumulative distribution of the defense year (smoothed over 2 years) of participants by categories (blue: category A-permanent, red: category B-post-docs, green: category C-out of astronomy, grey:unknown). Right:
Number of years spent as a post-doc by the participants (the darker part correspond to women, the lighter part correspond to men).}
\label{fig:duree}
\end{figure}

The poll was opened at this address 
\url{http://sondagedocteurastro.wikidot.com} and 200 replies were collected between February 
and june 2012. It was sent to the SF2A newsletter (1400 addresses), astronomy masters,
a French young searcher mailing list, and advertised in a ``Doctor in Astronomy'' group in Facebook.
Many questions were asked to the participants (year of the thesis, subject, job...) especially concerning the
utility of their studies in astronomy, and letting open the possibility to send comments. 
No field was compulsory.
While it was opened to every doctors in astronomy who obtained their degree in France,
the people who replied defended their thesis mainly between 2000 and 2010 (see Fig. \ref{fig:duree}). 
Older doctors may not be so interested in the destiny of their students, did not feel concerned, 
or are lazier. Many of the results are already available online (\url{http://sondagedocteurastro.wikidot.com/results}) and I will focus in the next section only on one important question.
  
\section{To be (a doctor) or not to be ? }

\begin{figure}[ht!]
 \centering
 \includegraphics[width=0.9\textwidth,clip]{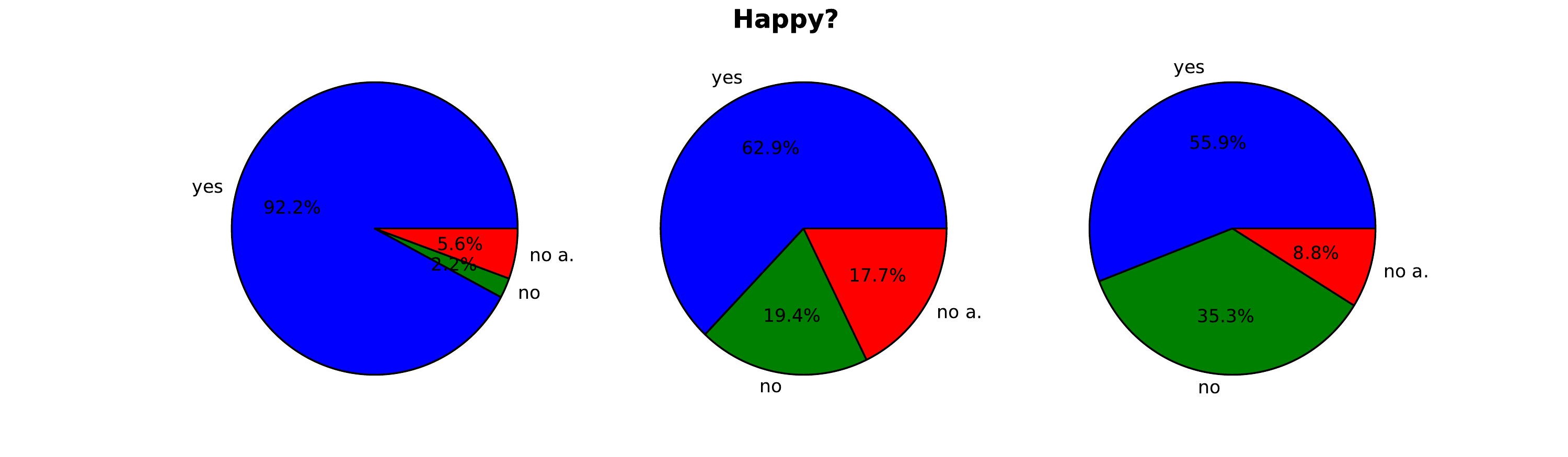}    \\   
 \includegraphics[width=0.9\textwidth,clip]{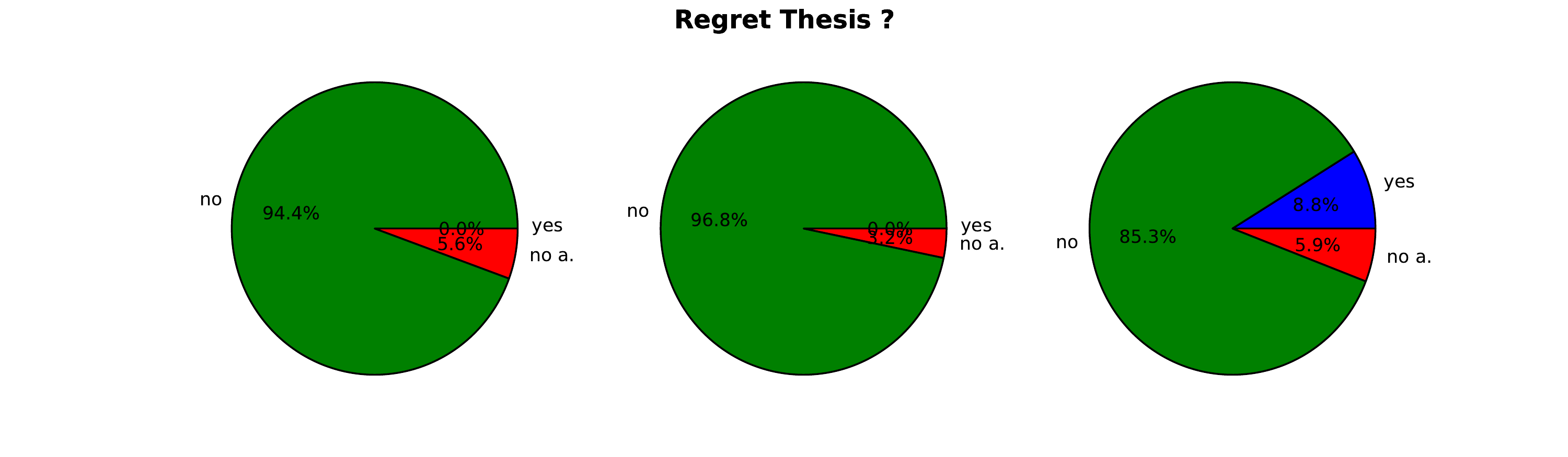}   \\
 \includegraphics[width=0.9\textwidth,clip]{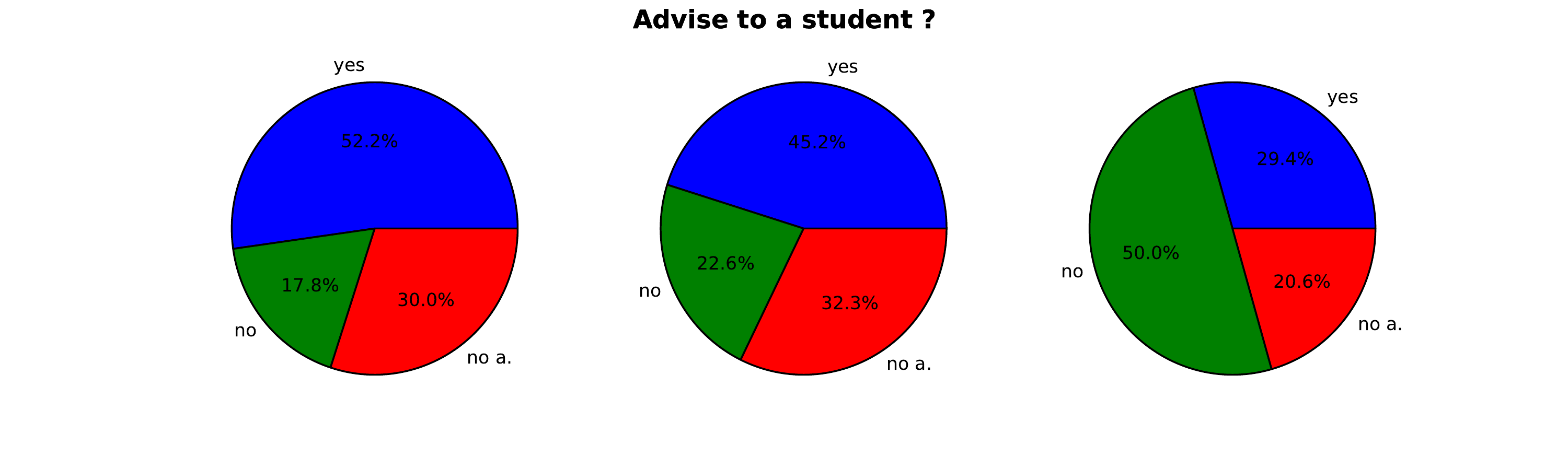}%
\vspace*{-0.8cm}  
  \caption{The top row shows the happiness of the poll participants concerning
their current situation, the middle row their reply to the question ``do you 
have regrets about having done a thesis in astronomy ?'', the bottom row their reply to
the question ``would you advice a student to start a thesis in astronomy ?''. From left 
to right : permanent staff, post-docs, people who left astronomy. ``no a.'' means ``no answer''.  }
  \label{fig:camembers}
\end{figure}

200 replies are too few to be complete or representative. Nevertheless, I tried to define three populations and analysed their respective replies : 
\begin{itemize}
\item A) permanent in astronomy, 
\item B) post-docs, 
\item C) people who left astronomy. 
\end{itemize}
While their relative numbers are not representative, each of this group is 
homogeneous enough to allow a comparison. Also, with an incomplete poll, ``success rates''
are meaningless, thus I will focus on the feelings of the participants 
(happiness concerning their situation, usefulness of their studies, advice to 
a student, see the distributions of the answers in Fig. \ref{fig:camembers}) 
and their commentaries.

Most of the participants are happy about their current
situation (we remind that this is not a complete survey). Most
have no regrets about having done a thesis (even among those with
no permanent positions, or  those who left astronomy). This by
itself is a nice message: a thesis is in general a good personal
experience, whatever happens (except in a few exceptional cases). However,
we should not be too happy, as only about 50 \% of \emph{permanent}
staff would advise a young student to start an astronomy thesis. This
number fall to 30 \% among those who left astronomy for which on the
contrary 50\% would not advise a student to start a thesis.
Moreover, the percentage of people suggesting to start
a thesis decreases with the thesis defense year, even when the study is
restricted to permanent staff : the situation is getting worse.

In my opinion, this negative message (that may push good students towards
other careers) is  directly  linked to the increasing 
pressure on jobs (and thus the increase of the number of thesis defended each year).  
The comments collected during the study allow us to understand even more why the
fraction not advising to start a thesis is so high. They stressed:
\begin{itemize}
\item[$\bullet$] the very bad preparation for a career outside
  academia. Some of my colleagues may say ``we do science,
  students are here to learn to do science''. While I respect this
opinion, I do not believe the argument is valid if we teach hordes
of students how to do science, while they will not do science in the 
future. If we want to perform ``training through research'' with large
number of students, then the one among them  ending up elsewhere must have gained
a useful \emph{formation} in the process. The alternative is of course to train
less students.

\item[$\bullet$] the bad connection between academia and the private
  sector (teachers in astronomy are unaware of the existence of local
  enterprises that may be interested in our formations).

\item[$\bullet$] the fact that having an astronomy degree is not 
recognized in the private sector, and simply outside astronomy.

\item[$\bullet$] the length of the the unstable life as a post-doc and
its negative effects on personal life.

\item[$\bullet$] the very large role of chance (see also the
  discussion in Mamon 2003), and especially the advantage for a
  student to have had a supervisor well-known in the community or
  participating to hiring committees. I remind that this is not a
  personal assertion but an opinion frequently stated in the replies
  to the poll (even from category A participants).  If the
  young doctors have this impression, it might be that it is partly
  true (\emph{``il n'y a pas de fum\'ee sans feu''}) or that some of the
  hiring process lack transparency.
\end{itemize}

\section{Some conclusions}

As members of the French astronomy community, we 
should be careful about the situation of young PhDs to prepare
a better future for our domain (and make it more 
attractive to \emph{good} students). 
At the present, I do not think there is even a
coherent vision at the national level of the
ensemble of thesis defended in astronomy each year in France.
A listing of all of those would be interesting 
and allow us to  follow more easily the becoming of PhDs and 
their career (in astronomy or elsewhere). May be the SF2A 
should perform this service for the community.

Concerning the number of PhDs, we need to  be logical and face the facts.
If we want to increase the number of PhD students well beyond the
number of opened jobs in astronomy each years (and this seems to be the
mainstream trend according to what I hear from some of my colleagues;
and the increasing funding possibilities -e.g. ERC, ANR-), then we
have to clearly state that we are developing a ``training through
research'' model.  
But \emph{saying it} is not \emph{doing it}. In order to adopt this
model we \emph{have to} make
sure that a PhD in astronomy is a real ``formation'', a positive
professional experience so that persons spending 3 years of their
life on an astronomical project have really earned 
something useful also outside the small world of research.
If we do not make this effort, it means that we are not forming people,
but just employing cheap manpower (I believe most astronomers 
have too much honnor to do such a thing).
How to do it is another question. We may need to 
establish stronger bridges with industries, to include in
our formations lectures useful also for non academic careers, 
or to push institutions to open more easily their doors to 
PhDs.
%
On the contrary, if we decide to be happy to form our students 
only to astronomy and that this is not a preparation
for anything else (an acceptable proposition), then we have to 
start lowering down the number of thesis in astronomy starting each year. 
The best way to do so is for each of us not to increase artificially 
the number of persons formed in astronomy  by using funding from e.g.
ANR, ERC grants (that did not exist a few years ago). 
Whatever is our choice (the debate is open), it is our collective responsibility to make one and
act in consequences.

\begin{acknowledgements}
I decided to write this proceeding by myself because it includes very personal
motivations and conclusions. However, this study would not have existed without the suggestions of S. Vives whom I thank deeply for motivating a wider work that I originally thought of. 
I also thank many people who encouraged me and made suggestions, citing a few of them: M.H. Aumeunier, M. Boquien, L. Ciesla, O. Cucciati, S. Heinis. I thank the SF2A council who fully supported this idea. I thank  finally all the persons who replied to the poll, and especially those who sent to me detailed comments, suggestions, mentions to other studies or even complementary data (especially G. Bonello, M. Boquien, G. Mamon, M. Vincendon).
\end{acknowledgements}


\begin{thebibliography}{}

\bibitem[Durette et al.(2012)]{durette} Durette, B.,  Fournier, M., Lafon, M., 2012, Comp\'etence et employabilit\'e des docteurs, \url{http://www.competences-docteurs.fr/} 

\bibitem[Mamon(2003)]{mamon} Mamon, G., 2003, The selection of tenured astronomers in France, arxiv:03033552

\bibitem[Metcalfe(2007)]{metcalfe} Metcalfe, T., 2007, The production rate and employment of PH.D. astronomers, arxiv:0712.2820

\bibitem[Seth et al.(2010)]{Seth} Seth, A., Ag{\"u}eros, M., 
Covey, K., et al.\ 2009, astro2010: The Astronomy and Astrophysics Decadal 
Survey, 2010, 51P 

\end{thebibliography}
\end{document}